\documentclass[aps,prl,showpacs,superscriptaddress,twocolumn]{revtex4-1}
\usepackage{times}
\usepackage{amsmath}
\usepackage{graphicx}
\usepackage{subfigure}
\usepackage{epsfig}
\usepackage{dcolumn}
\usepackage{bm}
\usepackage{lineno}
\usepackage{color}
\usepackage[english]{babel}

\newcommand{\xb}{X_b}

\newcommand{\lum}{{\cal L}}
\newcommand{\eff}{\varepsilon}
\newcommand{\BR}{{\cal B}}
\newcommand{\jpc}{J^{PC}}
\newcommand{\piz}{\pi^0}

\newcommand{\twog}{\gamma\gamma}
\newcommand{\ones}{\Upsilon(1S)}

\newcommand{\fives}{\Upsilon(10860)}
\newcommand{\chiz}{\chi_{b0}}
\newcommand{\chio}{\chi_{b1}}
\newcommand{\chit}{\chi_{b2}}
\newcommand{\chib}{\chi_{bJ}}
\newcommand{\omechiz}{\omega \chiz}
\newcommand{\omechio}{\omega \chio}
\newcommand{\omechit}{\omega \chit}
\newcommand{\omechib}{\omega \chib}

\newcommand{\EE}{e^+e^-}
\newcommand{\MM}{\mu^+\mu^-}
\newcommand{\LL}{\ell^+\ell^-}
\newcommand{\pp}{\pi^+\pi^-}
\newcommand{\ppp}{\pi^+\pi^-\pi^0}
\newcommand{\ra}{\rightarrow}
\newcommand{\beq}{\begin{equation}}
\newcommand{\eeq}{\end{equation}}
\newcommand{\bitm}{\begin{itemize}}
\newcommand{\eitm}{\end{itemize}}

\def\Journal#1#2#3#4{{#1} {\bf #2}, #3 (#4)}

\def\PLB{Phys. Lett. B}
\def\PRL{Phys. Rev. Lett.}
\def\PRD{Phys. Rev. D}

\def\EPJC{Eur. Phys. J. C}

\begin{document}

\preprint{} \preprint{\vbox{ \hbox{   }
                        \hbox{Intended for {\it Phys. Rev. Lett.}}
                        \hbox{Authors: X.~H.~He, C.~P.~Shen, C.~Z.~Yuan, Y.~Ban}
                        \hbox{Committee: S. Eidelman (chair), A. Bondar and R. Mussa}
}}

\title{\quad\\[0.5cm]
\boldmath
Observation of $e^+e^-  \to \pi^+ \pi^- \pi^0 \chi_{bJ}$  and search for $X_b \to \omega \Upsilon(1S)$
at $\sqrt{s}=10.867$ GeV}


\noaffiliation
\affiliation{University of the Basque Country UPV/EHU, 48080 Bilbao}
\affiliation{Beihang University, Beijing 100191}
\affiliation{University of Bonn, 53115 Bonn}
\affiliation{Budker Institute of Nuclear Physics SB RAS and Novosibirsk State University, Novosibirsk 630090}
\affiliation{Faculty of Mathematics and Physics, Charles University, 121 16 Prague}
\affiliation{Chonnam National University, Kwangju 660-701}
\affiliation{University of Cincinnati, Cincinnati, Ohio 45221}
\affiliation{Deutsches Elektronen--Synchrotron, 22607 Hamburg}
\affiliation{Justus-Liebig-Universit\"at Gie\ss{}en, 35392 Gie\ss{}en}
\affiliation{The Graduate University for Advanced Studies, Hayama 240-0193}
\affiliation{Gyeongsang National University, Chinju 660-701}
\affiliation{Hanyang University, Seoul 133-791}
\affiliation{University of Hawaii, Honolulu, Hawaii 96822}
\affiliation{High Energy Accelerator Research Organization (KEK), Tsukuba 305-0801}
\affiliation{IKERBASQUE, Basque Foundation for Science, 48011 Bilbao}
\affiliation{Indian Institute of Technology Bhubaneswar, Satya Nagar 751007}
\affiliation{Indian Institute of Technology Guwahati, Assam 781039}
\affiliation{Indian Institute of Technology Madras, Chennai 600036}
\affiliation{Institute of High Energy Physics, Chinese Academy of Sciences, Beijing 100049}
\affiliation{Institute of High Energy Physics, Vienna 1050}
\affiliation{Institute for High Energy Physics, Protvino 142281}
\affiliation{INFN - Sezione di Torino, 10125 Torino}
\affiliation{Institute for Theoretical and Experimental Physics, Moscow 117218}
\affiliation{J. Stefan Institute, 1000 Ljubljana}
\affiliation{Kanagawa University, Yokohama 221-8686}
\affiliation{Korea Institute of Science and Technology Information, Daejeon 305-806}
\affiliation{Korea University, Seoul 136-713}
\affiliation{Kyungpook National University, Daegu 702-701}
\affiliation{\'Ecole Polytechnique F\'ed\'erale de Lausanne (EPFL), Lausanne 1015}
\affiliation{Faculty of Mathematics and Physics, University of Ljubljana, 1000 Ljubljana}
\affiliation{Luther College, Decorah, Iowa 52101}
\affiliation{University of Maribor, 2000 Maribor}
\affiliation{Max-Planck-Institut f\"ur Physik, 80805 M\"unchen}
\affiliation{School of Physics, University of Melbourne, Victoria 3010}
\affiliation{Moscow Physical Engineering Institute, Moscow 115409}
\affiliation{Moscow Institute of Physics and Technology, Moscow Region 141700}
\affiliation{Graduate School of Science, Nagoya University, Nagoya 464-8602}
\affiliation{Kobayashi-Maskawa Institute, Nagoya University, Nagoya 464-8602}
\affiliation{Nara Women's University, Nara 630-8506}
\affiliation{National Central University, Chung-li 32054}
\affiliation{National United University, Miao Li 36003}
\affiliation{Department of Physics, National Taiwan University, Taipei 10617}
\affiliation{H. Niewodniczanski Institute of Nuclear Physics, Krakow 31-342}
\affiliation{Niigata University, Niigata 950-2181}
\affiliation{Osaka City University, Osaka 558-8585}
\affiliation{Pacific Northwest National Laboratory, Richland, Washington 99352}
\affiliation{Peking University, Beijing 100871}
\affiliation{University of Pittsburgh, Pittsburgh, Pennsylvania 15260}
\affiliation{University of Science and Technology of China, Hefei 230026}
\affiliation{Seoul National University, Seoul 151-742}
\affiliation{Soongsil University, Seoul 156-743}
\affiliation{Sungkyunkwan University, Suwon 440-746}
\affiliation{School of Physics, University of Sydney, NSW 2006}
\affiliation{Department of Physics, Faculty of Science, University of Tabuk, Tabuk 71451}
\affiliation{Tata Institute of Fundamental Research, Mumbai 400005}
\affiliation{Excellence Cluster Universe, Technische Universit\"at M\"unchen, 85748 Garching}
\affiliation{Toho University, Funabashi 274-8510}
\affiliation{Tohoku University, Sendai 980-8578}
\affiliation{Department of Physics, University of Tokyo, Tokyo 113-0033}
\affiliation{Tokyo Institute of Technology, Tokyo 152-8550}
\affiliation{Tokyo Metropolitan University, Tokyo 192-0397}
\affiliation{University of Torino, 10124 Torino}
\affiliation{CNP, Virginia Polytechnic Institute and State University, Blacksburg, Virginia 24061}
\affiliation{Wayne State University, Detroit, Michigan 48202}
\affiliation{Yamagata University, Yamagata 990-8560}
\affiliation{Yonsei University, Seoul 120-749}
  \author{X.~H.~He}\affiliation{Peking University, Beijing 100871} 
  \author{C.~P.~Shen}\affiliation{Beihang University, Beijing 100191} 
  \author{C.~Z.~Yuan}\affiliation{Institute of High Energy Physics, Chinese Academy of Sciences, Beijing 100049} 
  \author{Y.~Ban}\affiliation{Peking University, Beijing 100871} 
  \author{A.~Abdesselam}\affiliation{Department of Physics, Faculty of Science, University of Tabuk, Tabuk 71451} 
  \author{I.~Adachi}\affiliation{High Energy Accelerator Research Organization (KEK), Tsukuba 305-0801}\affiliation{The Graduate University for Advanced Studies, Hayama 240-0193} 
  \author{H.~Aihara}\affiliation{Department of Physics, University of Tokyo, Tokyo 113-0033} 
  \author{D.~M.~Asner}\affiliation{Pacific Northwest National Laboratory, Richland, Washington 99352} 
  \author{V.~Aulchenko}\affiliation{Budker Institute of Nuclear Physics SB RAS and Novosibirsk State University, Novosibirsk 630090} 
  \author{T.~Aushev}\affiliation{Institute for Theoretical and Experimental Physics, Moscow 117218} 
% ...
\collaboration{The Belle Collaboration}


\begin{abstract}
The $e^+e^- \ra \pi^+ \pi^- \pi^0 \chi_{bJ}$ ($J=0,~1,~2$) processes are studied
using a 118~fb$^{-1}$ data sample acquired with the Belle detector at a center-of-mass energy of 10.867 GeV.
Unambiguous $\pi^+ \pi^- \pi^0 \chi_{bJ}$ $(J=1,~2)$, $\omega \chi_{b1}$ signals
are observed, and indication for $\omega \chi_{b2}$
is seen, both for the first time, and the corresponding cross section
measurements are presented.
No significant $\pi^+\pi^-\pi^0\chi_{b0}$ or $\omega\chi_{b0}$ signals are observed and
90\% confidence level upper limits on the cross sections
for these two processes are obtained.
In the $\pi^+\pi^-\pi^0$ invariant mass spectrum,
significant non-$\omega$ signals are also observed.
We search for the $X(3872)$-like state (named $X_b$)
decaying into $\omega \Upsilon(1S)$;
no significant signal is observed with a mass between $10.55$
and $10.65$ GeV/$c^2$.

\end{abstract}


\maketitle


Investigation of hadronic transitions between heavy quarkonia is
a key source of information necessary for understanding Quantum
Chromodynamics (QCD).
Heavy quarkonium systems are in general nonrelativistic and hadronic transitions
for the lower-lying states have largely been successfully described
using the QCD multipole expansion model~\cite{qcd}.
New aspects of hadronic transitions between heavy quarkonia have been explored using
a data sample collected with Belle at the $\fives$ resonance peak. The
anomalously large width of the $\fives \ra \pp \Upsilon(mS)$ $(m
=1,~2,~3)$ and $\pp h_b(nP)$ $(n=1,~2)$ transitions ~\cite{kfchen} has been interpreted within
various QCD models~\cite{quarkonium} as either due to the rescattering of the $B$ mesons~\cite{explanation1}
or due to the existence of a tetraquark state, $Y_b$, with a mass close to that of the $\fives$
resonance~\cite{explanation2}.
A detailed analysis of the three-body
$\EE \to \pp \Upsilon(mS)$ and $\EE \to \pp h_b(nP)$ processes
reported by Belle~\cite{zb} revealed the presence of two
charged bottomonium-like states, denoted as $Z_b(10610)^{\pm}$
and $Z_b(10650)^{\pm}$. A similar
investigation of $\ppp$ hadronic transitions between the
$\fives$ and $\chi_{bJ}$  $(J=0,~1,~2)$ may offer additional
insight into strong interactions in heavy quarkonium systems.

The observation of the $X(3872)$~\cite{x3872} in 2003
revealed that the meson
spectroscopy is far more complicated than the naive expectation of the
quark model. It is therefore natural to search for
a similar state with $\jpc = 1^{++}$ (called $\xb$ hereafter)
in the bottomonium system~\cite{Xb1, Xb2}.
The search for $\xb$ supplies important information about the discrimination of a compact
multiquark configuration and a loosely bound hadronic molecule
configuration for the $X(3872)$.
The existence of the $\xb$ is predicted in both the
tetraquark model~\cite{tetra} and those involving a molecular
interpretation~\cite{mole, xb_decay, mole2}.
Recently, the CMS Collaboration
reported a null search for such a state in the $\pp\ones$ final state~\cite{cms_xb}.
However, unlike the $X(3872)$, whose decays exhibit large isospin violation,
the $\xb$ would decay preferably into $\ppp \ones$ rather than $\pp\ones$
if it exists~\cite{gang, xb_decay, xb_decay2,karliner}.

In this Letter, we study the $e^+ e^-\ra\ppp\chib$ ($J=0,~1,~2$) processes
 with subsequent $\chib \ra \gamma \ones$,
$\ones \ra \LL$ ($\ell=e~{\rm or}~ \mu$) decays.
As the $X(3872)$ was observed in $e^+e^- \to \gamma X(3872)$
at center-of-mass energies around 4.26~GeV~\cite{ablikim}, we also search for
an $X(3872)$-like state $\xb$ decaying to $\omega\ones$ with $\omega \to \ppp$
in $e^+e^- \to \gamma X_b$ at higher energies.
The results are based on a 118~fb$^{-1}$ data sample
collected with the Belle detector at $\sqrt{s}=10.867$ GeV.
The Belle detector ~\cite{belle} operates at
the KEKB asymmetric-energy $\EE$ collider~\cite{kekb}.

The {\sc evtgen}~\cite{evtgen} generator is used to simulate Monte Carlo
(MC) events.
For the two-body decays $\EE \ra \omechib$ and
$\EE \ra\gamma\xb$ at $\sqrt{s}=10.867$ GeV,
the angular distributions are generated using the formulae
in Ref.~\cite{gen_angle}.
The $\xb$ is assumed to have a mass of 10.6 GeV/$c^2$ and a negligible width in the MC simulation.
Other masses and widths are taken
from Ref.~\cite{pdg}.


For charged tracks, the impact parameters perpendicular to and along
the positron beam direction (the $z$ axis) with respect to the interaction point are required
to be less than 0.5~cm and 3.5~cm, respectively, and the transverse
momentum is restricted to be higher than 0.1~$\hbox{GeV}/c$.
A likelihood ${\mathcal{L}}_P$ for each charged track is obtained from
different detector subsystems for each particle hypothesis
$P$ $\in \{ e,\ \mu,\ \pi,\ K,\ p \}$.
Tracks with a likelihood ratio
$\mathcal{R}_K = \mathcal{L}_K/(\mathcal{L}_K+\mathcal{L}_\pi) < 0.4$ are identified as pions~\cite{pid}
with an efficiency of 96\%, while 4\% of kaons are misidentified as pions.
Similar likelihood ratios $\mathcal{R}_e$ and $\mathcal{R}_\mu$ are
defined for electron and muon identification~\cite{lid}.
The charged track
is accepted as an electron/positron if $\mathcal{R}_e > 0.01$
or as a muon if $\mathcal{R}_\mu > 0.1$.
The lepton pair identification efficiency is about
95\% for $\ones \ra \EE$ and 93\% for $\ones \ra \MM$.
Events with $\gamma$ conversion are removed by requiring
$\mathcal{R}_e<0.9$ for the $\pp$ candidate tracks.
Final state radiation and bremsstrahlung energy loss are recovered by adding
the four-momentum of photons detected within a 50 mrad cone around the
electron or positron flight direction in the $\EE$ invariant mass calculation.
The $\ones$ candidate is reconstructed from a pair of oppositely-charged leptons.

A neutral cluster in the electromagnetic calorimeter is
reconstructed as a photon if it does not match the extrapolated
position of any charged track and its energy is greater
than 50 MeV. To calibrate the photon energy resolution function,
three control channels $D^{\ast0}\to \gamma D(\to K^- \pi^+)$,
$\pi^0 \to \gamma \gamma$ and $\eta \to \gamma \gamma$ are used~\cite{calibration}.
A $\piz$ candidate is reconstructed from a pair of photons.
We require $M(\gamma \gamma)$ within $\pm$13 MeV/$c^{2}$ of the $\pi^0$ nominal mass as the signal region and
the non-$\piz$ backgrounds ($\piz$ sidebands) are defined as
0.08 GeV/$c^2 < M(\twog) < 0.115$ GeV/$c^{2}$ or 0.155 GeV/$c^2 < M(\twog) < 0.18$ GeV/$c^{2}$.

To improve the track momentum and photon energy resolutions and to reduce the
background, a five-constraint (5C) kinematic fit
is performed, where the  invariant mass of the two leptons
is constrained to
the $\ones$ nominal mass~\cite{pdg} and the energy and momentum of the
final-state system are
constrained to the initial $\EE$ center-of-mass system.
The $\chi^{2}_{{\rm 5C}}/{\rm dof}$ value is required
to be less than 5 for both $\ones\ra\ell^+ \ell^-$ modes with an efficiency of 85\%.
Here, ${\rm dof}=5$ is
the number of degrees of freedom.
This requirement removes events with one or
more additional or missing particles in the final states.
If there are multiple combinations for a candidate event,
the one with the smallest $\chi^2_{{\rm 5C}}/{\rm dof}$ is retained.

\begin{figure}[htbp]
\begin{center}
\includegraphics[width=8cm]{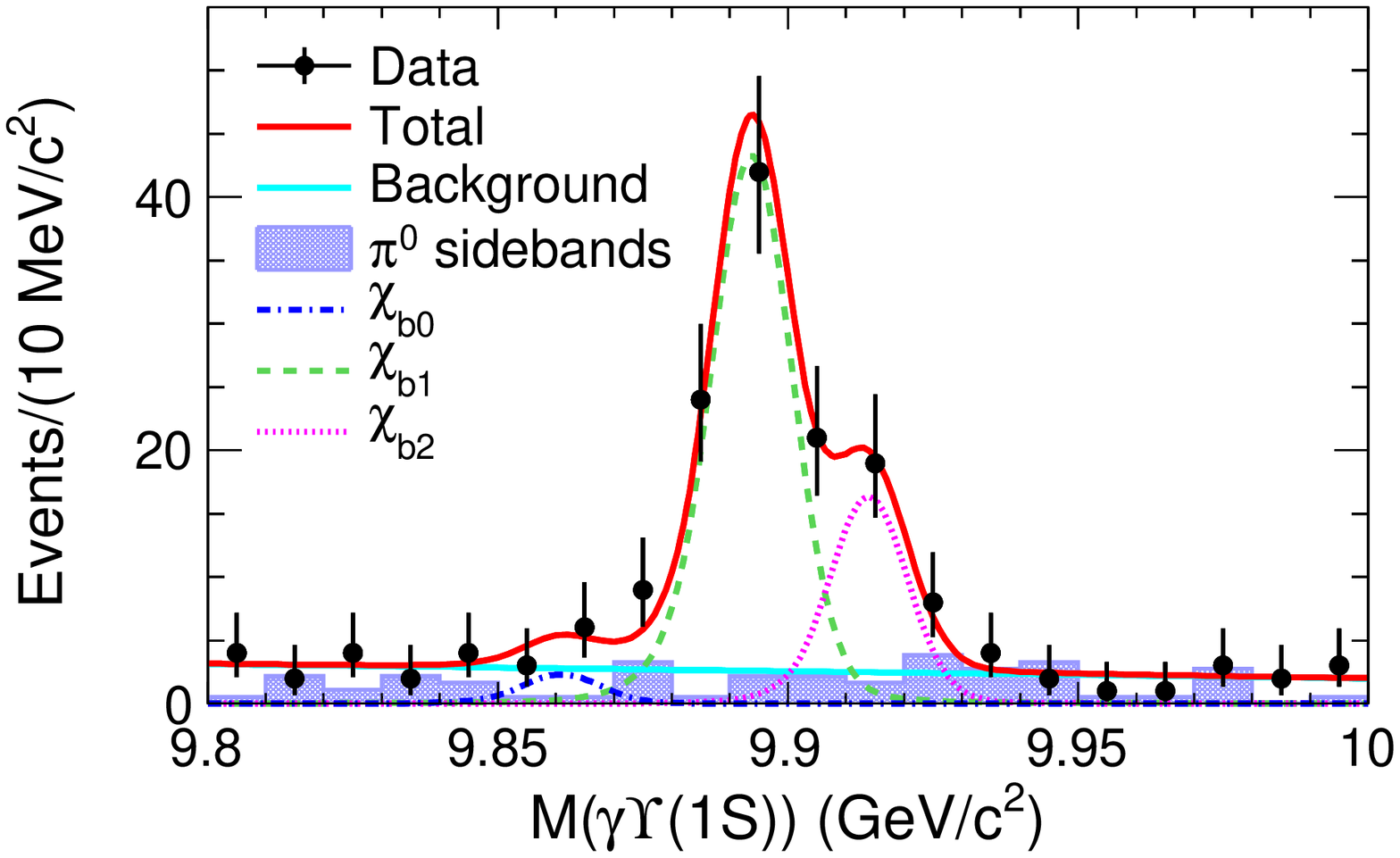}
\setlength{\abovecaptionskip}{1pt}
\caption{
The $\gamma \ones$ invariant mass distribution  for
selected $\EE \to\ppp \gamma \ones$ candidate events.
The shaded histogram is from
normalized $\piz$-sideband events. The fit to the $\gamma \ones$
invariant mass spectrum is described in the text.
The solid curves are the best fit for the total fit and background shape;
the dash-dotted, dashed and dotted curves represent the
$\chiz$, $\chio$ and $\chit$ signals, respectively.}
\label{mchi_3pi}
\end{center}
\end{figure}

The $\chib$ candidates are reconstructed from a candidate $\ones$ and a photon.
The $\gamma \ones$ invariant mass distribution after event selection
is shown in Fig.~\ref{mchi_3pi}, where the shaded histogram is from
the normalized non-$\piz$ background events.
Clear peaking signals in the $\chio$ and $\chit$ mass regions are observed.
We also examine the events in the $\chi^2_{5C}$
sidebands, defined as $15<\chi^2_{{\rm 5C}}/{\rm dof}<25$:
no $\chi_{bJ}$ peaks in the $M(\gamma \Upsilon(1S))$ distribution are found for
such events.

After the application of all of the selection requirements, the
remaining background comes mainly
from non-$\piz$ events that are represented by the $\pi^0$ sidebands
or possibly a subdominant non-$\chib$ background.
To probe for other peaking backgrounds,
a 89.4 fb$^{-1}$  continuum data sample collected at $\sqrt{s}=10.52$ GeV
and inclusive $\fives$ decays generated with {\sc
pythia}~\cite{pythia} with three times the luminosity of the data
are analyzed.
Moreover, MC samples of
$\fives \ra \eta\Upsilon(2S)$ $\to$ $\gamma\gamma\pp\ones$, $\fives$ $\to$ $\pp\Upsilon(2S)$$\to$ $\pp\piz\piz\ones$,
$\fives$ $\ra$ $\piz\piz\Upsilon(2S)$ $\ra$ $\piz\piz\pp\ones$,
$\fives$ $\ra$ $\pp\Upsilon(2S)$ $\ra$ $\pp\gamma\chio$ $\ra$ $\pp\gamma\gamma\ones$
and $\fives$ $\ra$ $\pp\Upsilon(1D)$ $\ra$ $\pp\gamma\chio$ $\ra$ $\pp\gamma\gamma\ones$
are generated and analyzed: no structures in the $\gamma\ones$
invariant mass spectrum are seen in these samples after applying all of the selection criteria.

An unbinned extended maximum likelihood fit is applied to the
$\gamma \ones$ mass spectrum  with Crystal Ball functions~\cite{cb}
(parameters being fixed to
the values from the fits to $\gamma \ones$ mass spectra from MC
signal samples) as $\chib$ signal
shapes and a first-order polynomial function as a background shape.
Figure~\ref{mchi_3pi} shows the fit results.

The statistical significance of the signal is estimated from the difference of the
logarithmic likelihoods~\cite{significance}, $-2\ln(L_0/L_{\rm max})$, where $L_0$ and $L_{\rm
max}$ are the likelihoods of the fits without and with a signal
component, respectively, taking the
difference in the number of degrees of freedom ($\Delta\hbox{dof}=1$) in the fits into account.
The signal significances of $\chio$ and $\chit$ are 12$\sigma$ and 5.9$\sigma$
with systematic uncertainties included, while for the $\chiz$
the signal significance is only 1.0$\sigma$. The fit results including
the signal yield, detection efficiency, signal significance,
and the calculated Born cross section for each mode are summarized
in Table~\ref{result}.
The Born cross section is calculated using
$\sigma_B=N\cdot|1-\Pi|^2 /[\lum\cdot\BR_{\rm int}\cdot\epsilon\cdot(1+\delta)]$,
where $N$ is the signal yield, $\lum$ is the integrated luminosity, $\BR_{\rm int}$
is the product of the branching fractions of the intermediate states to
the reconstructed final states, $\epsilon$ is the corresponding detection
efficiency, $1+\delta$ is the radiative correction factor and $|1-\Pi|^2$ is the vacuum polarization
factor. In the MC simulation, trigger efficiency is included,
and initial state radiation is taken into account by
assuming the cross sections follow the $\fives$ line shape
with a zero non-resonant contribution~\cite{pdg}.
The radiative correction factor $1+\delta$ is $0.65\pm0.05$ calculated
using the formulae in Ref.~\cite{rc}; the value of $|1-\Pi(s)|^2$ is 0.929~\cite{vac}.
The calculated branching fraction $\BR$ for each mode is also shown in Table~\ref{result},
where the total number of $\fives$
events is (4.02 $\pm$ 0.20)$ \times 10^{7}$
using $\sigma_{b \bar{b}}\equiv \sigma(e^+e^- \to b \bar{b})=(0.340\pm0.016)$ nb~\cite{5snum} and assuming
all the $b \bar{b}$ events are from $\fives$ resonance
decays~\cite{brstate}.

We determine a Bayesian 90\% confidence level (C.L.) upper limit
on the number of $\chiz$ signal events ($N_{\rm sig}$)
by finding the value $N^{\rm UP}_{\rm sig}$ such that
$
\int_{0}^{N^{\rm UP}_{\rm sig}} \mathcal{L} dN_{\rm sig}/\int_{0}^{\infty} \mathcal{L} dN_{\rm sig}=0.90,
$
where $N_{\rm sig}$ is the number of $\chiz$ signal events and $\mathcal{L}$ is
the value of the likelihood as a function of $N_{\rm sig}$.
To take into account the systematic uncertainty, the above likelihood
is convolved with a Gaussian function whose width equals the total
systematic uncertainty. The upper limit on the number of
$\chiz$ signal events is 13.6 at 90\% C.L.

\begin{figure}[htbp]
\begin{center}
\subfigure{
\includegraphics[width=4.3cm]{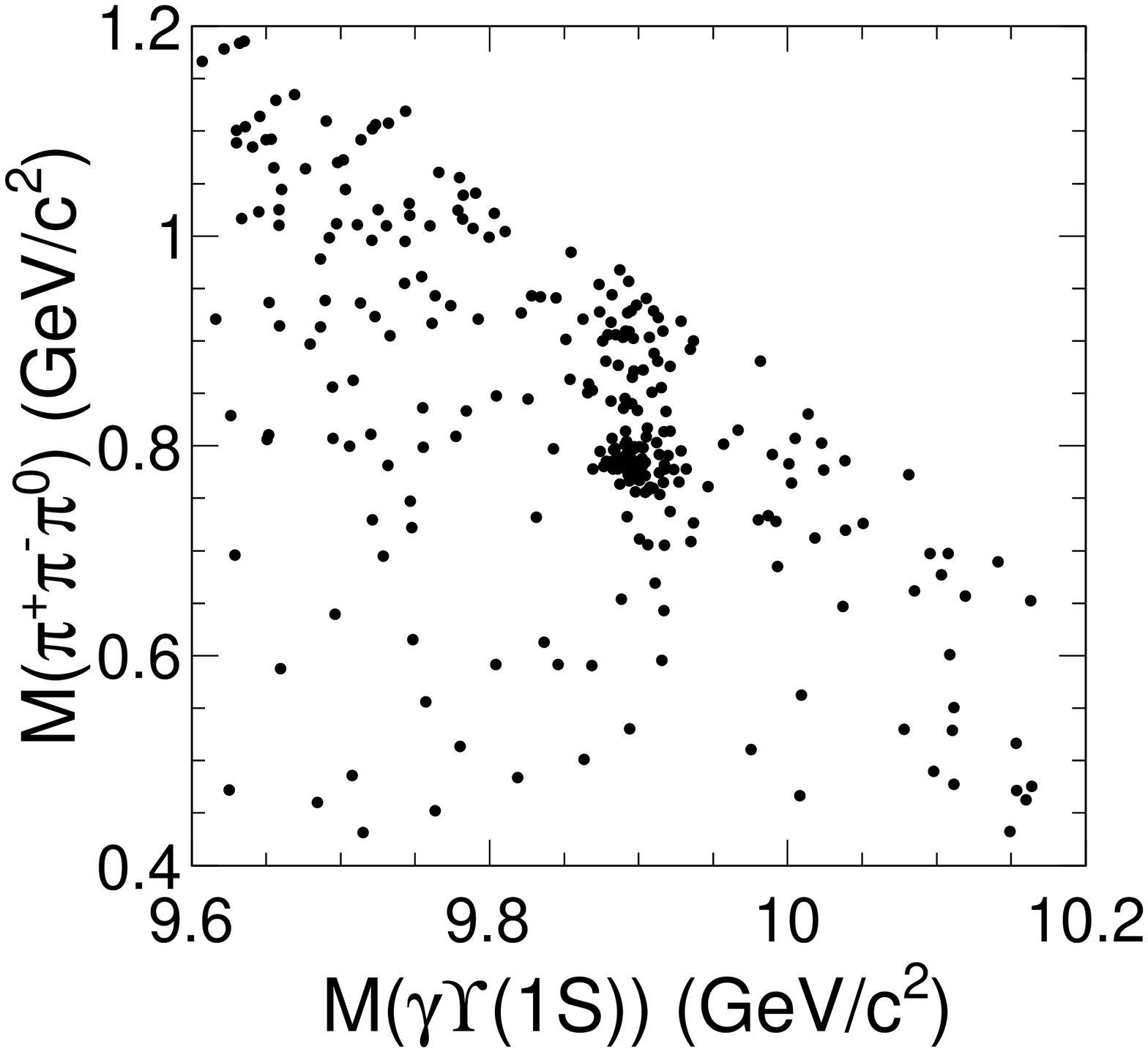}
\put(-35,85){\bf \large{(a)}}
} \hspace{-0.3cm}
\subfigure{
\includegraphics[width=4.3cm]{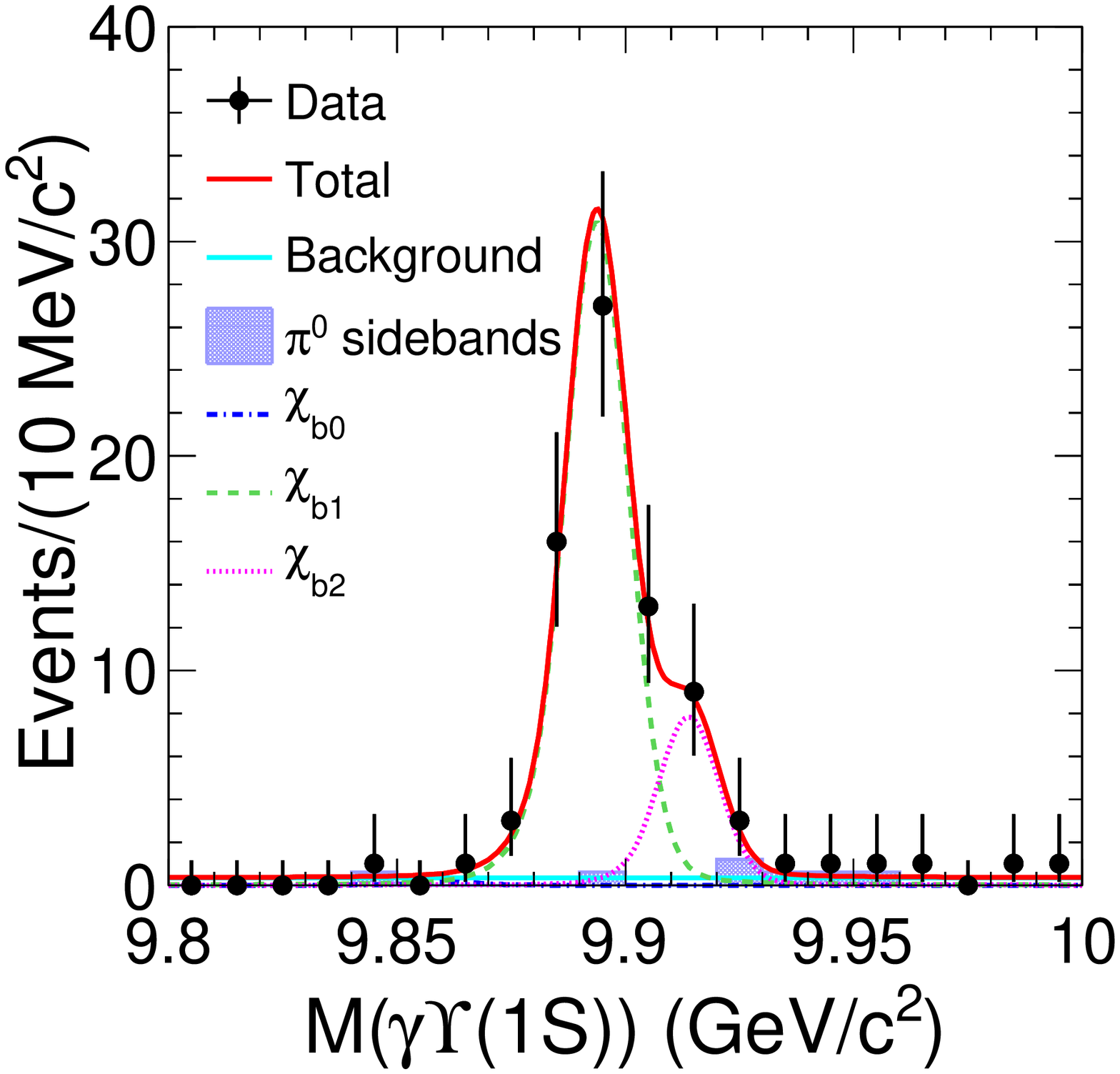}
\put(-35,85){\bf \large{(c)}}
}
\subfigure{
\includegraphics[width=4.3cm]{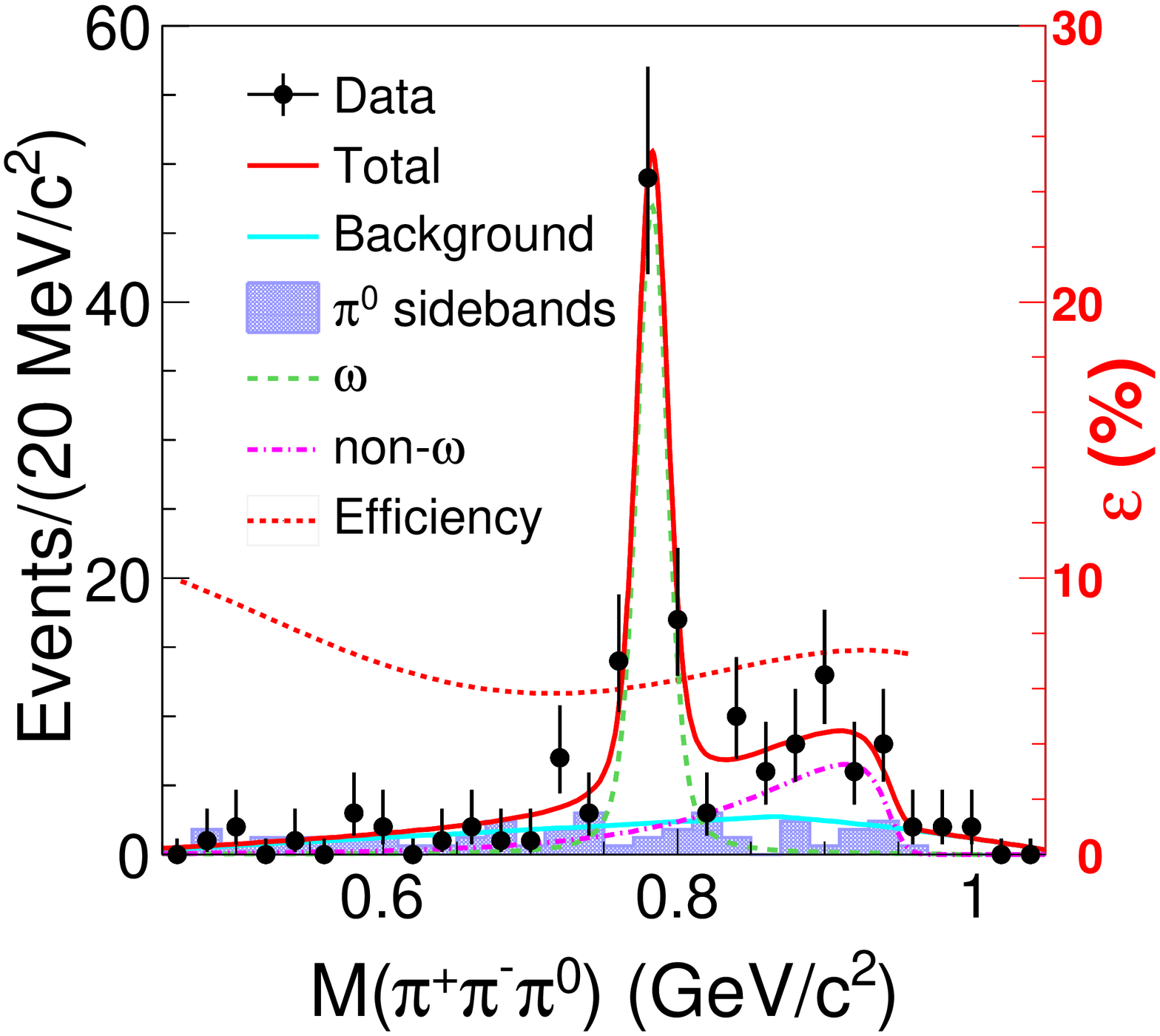}
\put(-35,85){\bf \large{(b)}}
}\hspace{-0.21cm}
\subfigure{
\includegraphics[width=4.3cm]{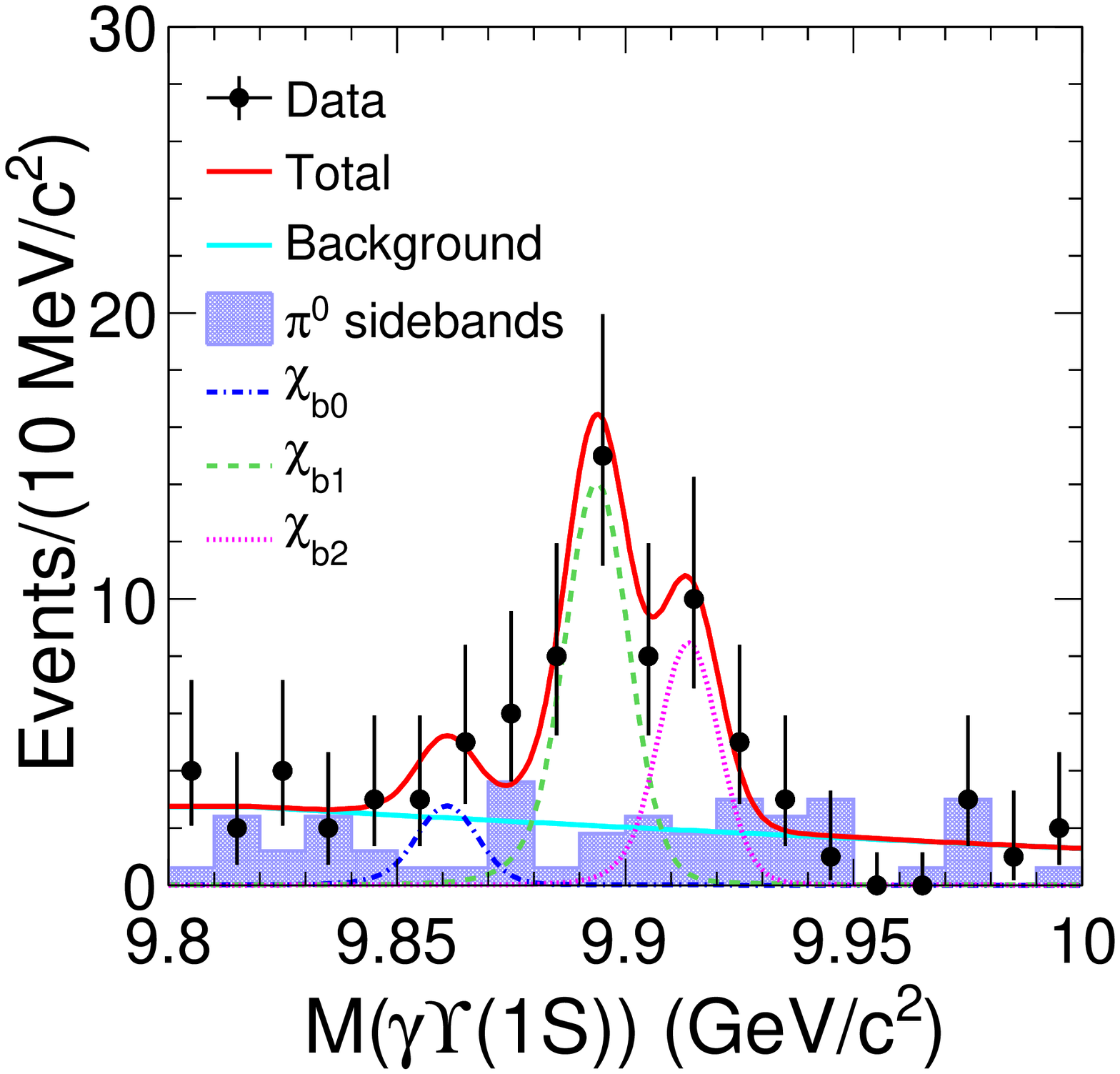}
\put(-35,85){\bf \large{(d)}}
}
\setlength{\abovecaptionskip}{1pt}
\caption{
(a) The scatter plot of $M(\ppp)$ versus $M(\gamma\ones)$
for selected $\EE \to\ppp \gamma \ones$ candidate events; and (b) the projections to $M(\ppp)$
for 9.8 GeV/$c^2$ $<$$M(\gamma \Upsilon(1S))$$<$10 GeV/$c^2$, where the dashed and dash-dotted
curves represent the $\omega$ and $(\ppp)_{\mbox{non}-\omega}$ events;
the dotted curve shows the efficiency dependence on $M(\ppp)$. Projections of $M(\gamma \ones)$
(c) in the $\omega$ signal region and (d) outside of $\omega$ signal region,
where the dash-dotted, dashed and dotted curves represent the
$\chiz$, $\chio$ and $\chit$ signals, respectively.
The solid curves are the best fit for the total signal and background shapes.
The shaded histograms are from the normalized $\piz$ sideband
events.
}
\label{2dfit}
\end{center}
\end{figure}

Figure~\ref{2dfit}(a) shows the scatter plot of $M(\ppp$) versus $M(\gamma\ones$).
Besides the clear $\omega$ signal in the $\chib$ mass region,
there is an obvious accumulation of events above the $\omega$ mass region.
Hereinafter, we denote these events as $(\ppp)_{\mbox{non}-\omega}$ events.

An unbinned two-dimensional (2D) extended maximum likelihood fit
to the $\ppp$ versus $\gamma\ones$ mass distributions
is applied to extract the  $\omega \chib$ and
$(\ppp)_{\mbox{non}-\omega} \chib$ yields.
In this fit, Crystal Ball functions
(parameters being fixed to
the values from the fits to $\gamma \ones$ mass spectra from MC
signal samples) are used for the $\chib$ signal
shapes, a Breit-Wigner function and an Argus function~\cite{argus}
(both are convolved with a Gaussian resolution function)
represent the $\omega$ and $(\ppp)_{\mbox{non}-\omega}$ shapes, respectively,
and a linear function is used for the backgrounds.
The Gaussian resolution function is obtained from MC simulation.

Figures~\ref{2dfit}(b-d) show the
$\ppp$ mass projection for 9.8 GeV/$c^2$ $<M(\gamma \Upsilon(1S))<10$ GeV/$c^2$, and the $\gamma \ones$ mass projection
within and outside the $\omega$ signal region (0.753 GeV/$c^2$ $<M(\ppp)<$ 0.813 GeV/$c^2$), where the shaded histograms are from
the normalized $\piz$ sideband events.
Clear $\chi_{b1}$ and $\chi_{b2}$ signals can be seen in the $\gamma \ones$
invariant mass spectrum,
while no excess of $\chi_{b0}$ events above expected backgrounds is observed.
The fit results with the calculated Born cross sections and branching fractions are summarized
in Table~\ref{result}.

\begin{table*}[htbp]
\caption{\label{result} Fitted signal yield, signal significance ($\Sigma$),
detection efficiency ($\eff$), Born cross section ($\sigma_B$), branching fraction ($\BR$) and
relative systematic uncertainty ($\sigma_{\rm sys}^{(1)}$ for Born cross section and $\sigma_{\rm sys}^{(2)}$ for branching fraction).
The upper limits are given at 90\% C.L. for the decay modes with
a signal significance of less than 3$\sigma$.
}

\begin{center}
\tabcolsep=5pt
\begin{tabular}{l|ccccccc}
\hline\hline
Mode    & Yield   & $\Sigma$  $(\sigma)$   & $\eff$ (\%)  & $\sigma_{B}$ (pb)  & $\BR$ ($10^{-3}$)  & $\sigma^{(1)}_{\rm sys}$ (\%) & $\sigma^{(2)}_{\rm sys}$ (\%) \\
\hline
$\ppp\chiz$    & $<13.6$           & 1.0   & 6.43     & $<$ 3.1         & $<$ 6.3          & 25 & 24         \\
$\ppp\chio$    & 80.1 $\pm$ 9.9    & 12    & 6.61     & 0.90 $\pm$ 0.11 $\pm$ 0.13  & 1.85 $\pm$ 0.23 $\pm$ 0.23   & 14 & 12  \\
$\ppp\chit$    & 28.6 $\pm$ 6.5    & 5.9   & 6.65     & 0.57 $\pm$ 0.13 $\pm$ 0.08  & 1.17 $\pm$ 0.27 $\pm$ 0.14   & 14 & 12  \\
\hline
$\omechiz$     & $<$ 7.5           & 0.5   & 6.35     & $<$ 1.9          & $<$ 3.9           & 29 & 28         \\
$\omechio$     & 59.9 $\pm$ 8.3    & 12    & 6.53     & 0.76 $\pm$ 0.11 $\pm$ 0.11  & 1.57 $\pm$ 0.22 $\pm$ 0.21   & 14 & 13 \\
$\omechit$     & $12.9\pm 4.8$     & 3.5   & 6.56     & 0.29 $\pm$ 0.11 $\pm$ 0.08  & 0.60 $\pm$ 0.23 $\pm$ 0.15   & 26 & 25 \\
\hline
$(\ppp)_{\mbox{non}-\omega}\chiz$     & $<$ 10.7          & 0.4   & 6.68     & $<$ 2.3          & $<$ 4.8           & 41 & 41      \\
$(\ppp)_{\mbox{non}-\omega}\chio$     & 23.6 $\pm$ 6.4    & 4.9   & 6.88     & 0.25 $\pm$ 0.07 $\pm$ 0.06  & 0.52 $\pm$ 0.15 $\pm$ 0.11   & 21 & 20 \\
$(\ppp)_{\mbox{non}-\omega}\chit$     & $15.6\pm 5.4$     & 3.1   & 6.91     & 0.30 $\pm$ 0.11 $\pm$ 0.14  & 0.61 $\pm$ 0.22 $\pm$ 0.28   &45 & 45 \\
\hline\hline
\end{tabular}
\end{center}
\end{table*}

There are several sources of systematic errors for the
cross section and branching fraction measurements. Tracking
efficiency uncertainties are estimated to be 1.0\% per pion track
and 0.35\% per lepton track, which are fully correlated in the momentum and
angle regions of interest for signal events.
The uncertainty due to particle identification
efficiency is 1.3\% for each pion and 1.6\%
for each lepton, respectively. The uncertainty in
the calibration of the photon energy resolution
is less than 1.1\% by checking the difference
with and without the calibration.
The uncertainty in
selecting $\pi^0$ candidates is estimated by comparing control
samples of $\eta \to \pi^0 \pi^0 \pi^0$ and $\eta \to \pi^+ \pi^- \pi^0$
decays in data and amounts to 2.2\%.
The uncertainty due to the 5C kinematic fit is 4.2\% obtained by comparing
the final results with or without using this fit.
A 3.0\% systematic error is assigned to the trigger uncertainty.
Errors on the branching fractions
of the intermediate states are taken from Ref.~\cite{pdg}.
For the cross section measurement, the uncertainty of the
total luminosity is 1.4\%. For the branching fraction
measurement, the uncertainty on the total number of $\fives$ events is 4.9\%, which
incorporates the uncertainty of the cross section $\sigma(\EE \to b\bar{b})$ (4.7\%)~\cite{5snum}.
The uncertainty on the radiative correction factor is 7.7\%
due to the uncertainties of the $\fives$ resonant parameters.
The uncertainty due to limited MC statistics is at
most 1.0\%.
We estimate the systematic
errors associated with the fitting procedure by changing
the order of the background polynomial and the range
of the fit, and comparing the fit results without a $\chiz$ component.
Finally, the uncertainties due to the fitting procedure are 3.9\%, 1.6\%, 3.2\%
for $\ppp \chi_{bJ}$ $J=0$, 1, 2, respectively.
For the $\omega \chi_{bJ}$ processes, the uncertainties in the yields of $\chib$ events due
to the 2D fit model are estimated.
We modify the background shape
to a constant or a second-order polynomial and
the parametrization description for the $(\ppp)_{\mbox{non}-\omega}$ events
to a free Breit-Wigner function to check the results stability with respect
to the fit model. The maximum differences compared with the nominal results are taken as the
systematic uncertainties and are 15.8\%, 4.4\% and 21.7\%, for $\omechib$,
and 32.8\%, 14.1\% and 42.3\%, for $(\ppp)_{\mbox{non}-\omega} \chib$,
$J=0$, 1, and 2, respectively.
For $(\ppp)_{\mbox{non}-\omega} \chib$, an uncertainty due to the unknown spin-parity
of the $(\pi^+\pi^-\pi^0)_{\mbox{non}-\omega}$ system (6.0\%) is also included.
Assuming all the sources are independent
and adding them in quadrature, the final total systematic uncertainties
for the studied modes are
summarized in Table~\ref{result}.

\begin{figure}[htbp]
\begin{center}
\includegraphics[width=8cm]{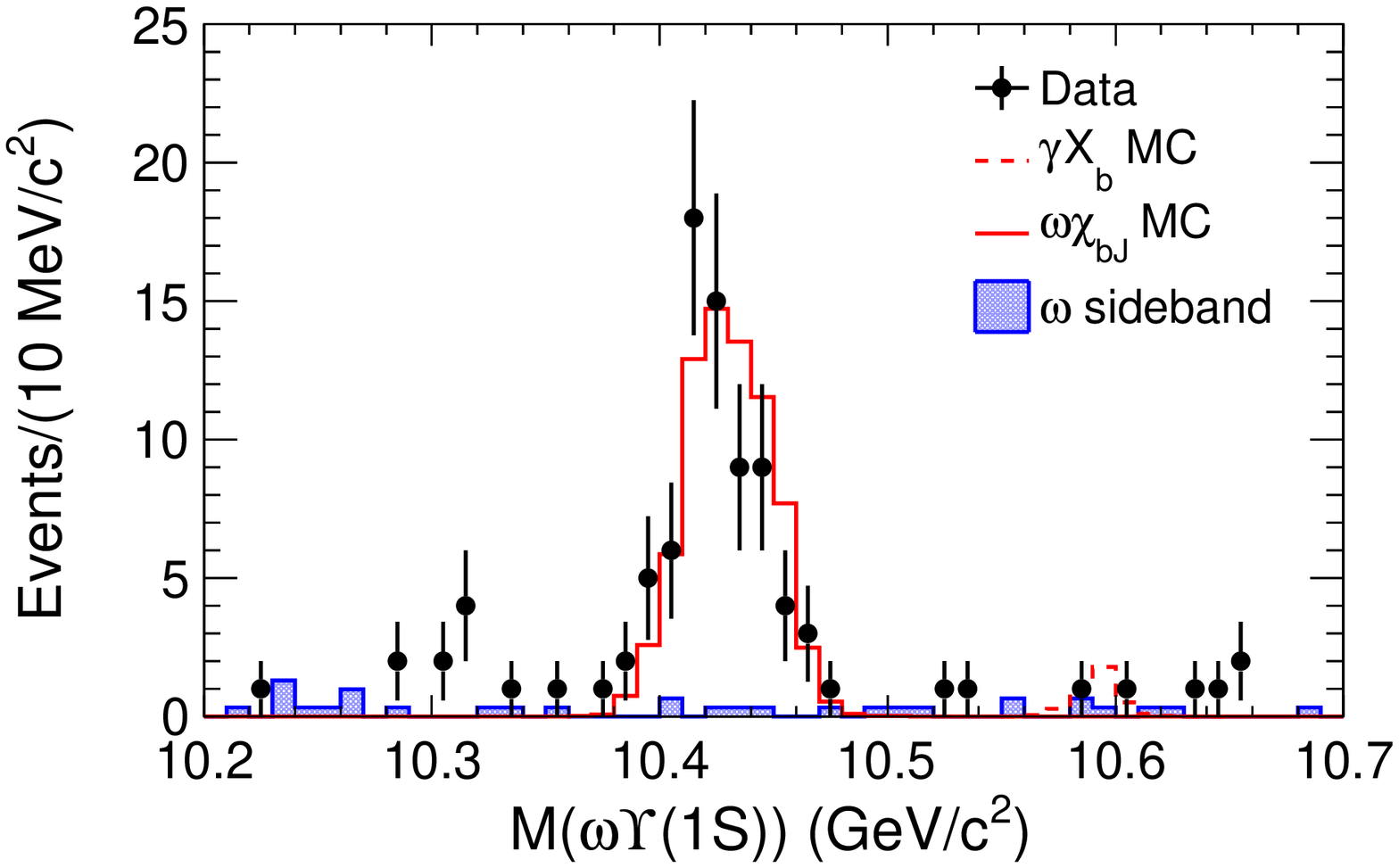}
\setlength{\abovecaptionskip}{1pt}
\caption{The $\omega\ones$ invariant mass distribution.
The dots with error bars are from data,
the solid histogram is from the normalized
contribution of $e^+e^- \to \omega \chi_{bJ}$ ($J=0,~1,~2$)
from MC simulation and the shaded histogram is from
normalized $\omega$ mass sideband events. The dashed histogram is from the MC signal sample
$\EE \ra \gamma \xb \ra \gamma \omega \ones\to \gamma \ppp \LL$
at $\sqrt{s}=10.867$ GeV with $\xb$ mass fixed at 10.6 GeV/$c^2$
and yield fixed at the upper limit at 90\% C.L.}
\label{mxb}
\end{center}
\end{figure}

We search for the $X(3872)$-like state $X_b$ in the process
$\EE \ra\gamma\xb$ with $\xb\ra\omega\ones$ at $\sqrt{s}=10.867$ GeV.
The selection criteria are the same as in $\EE\ra\ppp \chi_{bJ}$.
Figure~\ref{mxb} shows the $\omega\ones$ invariant mass distribution
with the requirement of $M(\pi^+ \pi^-\pi^0)$ within the $\omega$ signal
region; we search for
the $X_b$ from 10.55 to 10.65 GeV/$c^2$.
The dots with
error bars are from data,
the solid histogram is from the normalized
contribution of $e^+e^- \to \omega \chi_{bJ}$ ($J=0,~1,~2$)
and the shaded histogram is from the normalized
$\omega$ mass sideband, defined as 0.54 GeV/$c^2$ $<$ $M(\ppp)$ $<$ 0.72 GeV/$c^2$. No obvious $\xb$ signal is observed
after applying all the event selection criteria.

An unbinned extended maximum likelihood fit to the $\omega\ones$ mass
distribution is applied, where the signal shape is obtained from MC
simulation and the background is parameterized as a
first-order polynomial. From the fit, we obtain $-0.4\pm2.0$
$X_b$ signal events with a mass at $10.6$ GeV/$c^2$.
The upper limit on the yield of the $X_b$ signal events is
4.0 at 90\% C.L. with systematic uncertainty included. The dashed histogram in
Fig.~\ref{mxb} shows the upper limit on the yield of
$X_b$ signal events.

With the detection efficiency of 8.1\% and
assuming that the observed signals come from
$\fives$ decays, we obtain the product branching fraction
$\BR(\fives\ra\gamma\xb)\BR(\xb\ra\omega\ones)< 2.9\times10^{-5}$ at 90\% C.L.
The systematic uncertainties on the above branching fraction measurement
are almost the same as in $\EE \ra\omechib$, except for the fit uncertainty (29\%) and
total error on the branching fractions of the intermediate states (3.2\%).
Assuming all the sources are independent and adding them in quadrature,
we obtain a total systematic uncertainty of 31\%.
Using the aforementioned method, 90\% confidence level upper limits on the product branching fraction
$\BR(\fives\ra\gamma\xb)\BR(\xb\ra\omega\ones)$ vary smoothly
from $2.6\times10^{-5}$ to $3.8\times10^{-5}$ between $10.55$ and $10.65$ GeV/$c^2$.

In summary, using the 118~fb$^{-1}$ $\fives$ data sample
collected with Belle,
the processes $\EE \ra \ppp \chib$
and $\omechib$ ($J=0,~1,~2$) are
studied. We observe clear $\ppp \chio$ and $\ppp\chit$ signals,
while no significant $\ppp \chiz$ signal is found.
In the $\pi^+\pi^-\pi^0$ invariant mass spectrum, besides a clear $\omega$ signal,
significant non-$\omega$ signals are also observed.
The $\omechio$ signal and indication for $\omechit$ are found,
while no significant signal of $\omechiz$ can be seen.
All the results are summarized in Table~\ref{result}.
The measured branching fractions of $\fives \to \pi^+ \pi^- \pi^0 \chi_{b1}$
and $\pi^+ \pi^- \pi^0 \chi_{b2}$ are large and at the same order as the processes
$\fives \to \pp \Upsilon(mS)$ $(m
=1,~2,~3)$~\cite{kfchen}.
This is the first observation of hadronic transitions between the
$\fives$ and $\chi_{b1, b2}$ bottomonium states that provides important
information for understanding QCD dynamics.
The measured ratio of the branching fractions of $\fives$ decays or the cross sections of $\EE$
to $\omega\chi_{b2}$ and $\omega\chi_{b1}$
is $0.38\pm0.16(\rm stat.)\pm 0.09(\rm syst.)$,
where the common systematic uncertainties cancel.
It is significantly lower than the expectation of 1.57
from the heavy quark symmetry~\cite{hqs, 22}.
For $(\pi^+\pi^-\pi^0)_{\mbox{non}-\omega}$ events, such ratio
is $1.20\pm0.55(\rm stat.)\pm 0.65(\rm syst.)$.
We also search for the $X(3872)$-like state, $\xb$ with a
hidden $b\bar{b}$ component decaying into $\omega \ones$,
in $\fives$ radiative decay. No significant signal is observed for
such a state with mass between $10.55$ and $10.65$ GeV/$c^2$.


We thank the KEKB group for excellent operation of the
accelerator; the KEK cryogenics group for efficient solenoid
operations; and the KEK computer group, the NII, and
PNNL/EMSL for valuable computing and SINET4 network support.
We acknowledge support from MEXT, JSPS and Nagoya's TLPRC (Japan);
ARC and DIISR (Australia); FWF (Austria);
NSFC, the Fundamental Research Funds for the Central Universities YWF-14-WLXY-013
and CAS center for Excellence in Particle Physics (China); MSMT (Czechia);
CZF, DFG, and VS (Germany); DST (India); INFN (Italy);
MOE, MSIP, NRF, GSDC of KISTI, BK21Plus, and WCU (Korea);
MNiSW and NCN (Poland);
MES, RFAAE and RFBR grant 14-02-01220 (Russia);
ARRS (Slovenia);
IKERBASQUE and UPV/EHU (Spain);
SNSF (Switzerland); NSC and MOE (Taiwan); and DOE and NSF (USA).

{\it Note added}.-- After preliminary results were reported at the
international conferences, a few
theoretical models have been developed to interpret the data:
the possible cascade process
$\fives \to \pi Z_b \to \pi \rho \chi_{b}$ in
$(\pi^+\pi^-\pi^0)_{\mbox{non}-\omega} \chi_b$ events~\cite{11};
a molecular component in $\fives$ wave-function~\cite{11} or
an S- and D-wave mixing for the observed heavy quark symmetry
violation between $\omega \chi_{b1}$ and $\omega \chi_{b2}$~\cite{22};
and hadronic loop effect
for the large branching fractions measured~\cite{33}.



\begin{thebibliography}{**}
\bibitem{brstate} In case there is a non-$\fives$ resonance contribution in the
$b\bar{b}$ cross section, the definition of the branching fraction
$\BR$ is equivalent to the ratio of the two cross sections, i.e.,
$\BR=\sigma_{\rm vis}/\sigma_{b \bar{b}}=N/(\lum\cdot\BR_{\rm int}\cdot\eff\cdot\sigma_{b \bar{b}})$,
where $\sigma_{\rm vis}(=N/(\lum\cdot \BR_{\rm int}\cdot \eff))$ is the visible cross section.
\end{thebibliography}
\end{document}